# Online information on medical cannabis may rise unrealistic expectations and downplay potential side effects

Arthur Cassa Macedo[1], André Oliveira Vilela de Faria[1], Isabella Bizzi[2], Fabrício A. Moreira[1], Alessandro Colasanti[3] and Pietro Ghezzi[3]

[1] Faculdade de Medicina, Universidade Federal de Minas Gerais, Belo Horizonte, Brazil

[2] Universidade Federal do Rio Grande do Sul, Porto Alegre, Brazil

[3] Brighton & Sussex Medical School, Brighton, United Kingdom

**Abstract**

There is a growing literature on the potential medical uses of *Cannabis sativa* and cannabinoid compounds. Although these have only been approved by regulatory agencies for few indications, there is a hype about their possible benefits in a variety of conditions and a large market in the wellness industry. As in many cases patients search for information on cannabis products online, we have analyzed the information on medical cannabis available on the Internet. Analyzing 176 webpages returned by a search engine, we found that more than half of them were news websites. Pain, epilepsy and multiple sclerosis were the most frequently therapeutic areas mentioned by the webpages, which did not always match those for which there is regulatory approval. Information was also incomplete, with only 22% of the webpages mentioning potential side effects. Health portal websites provided the most complete information. On average, 80% of webpages had a neutral stance on the potential benefits of medical cannabis, with commercial websites having more frequently a positive stance (67%). We conclude that the information that can be found online could raise unrealistic expectations regarding therapeutic areas for which science-based evidence is often still weak.

# 1. Introduction

There is a growing literature on the potential medical uses of *Cannabis sativa* and its derived molecules, termed cannabinoids, including $\Delta^9$-tetrahydrocannabinol (THC) and cannabidiol (CBD) (World Health Organization Expert Committee on Drug Dependence, 2018). The pharmaceutical preparations available comprise plant-derived THC, synthetic THC (dronabinol), the synthetic cannabinoid nabilone, plant-derived CBD and a combination of plant-derived THC and CBD. Despite the recent explosion in interest, and suggestions that earliest medical use of cannabis products could be tracked back to ancient times (Zias, et al., 1993), at present only very few cannabis-based medicines have been approved for clinical use. In some countries, a plant extract containing a combination of approximately equal parts of THC and CBD is in clinical use for the alleviating the symptoms of multiple sclerosis. Moreover, a CBD-containing extract has been approved for the treatment of refractory epileptic syndromes, such as Lennox-Gastaut syndrome or Dravet syndrome, and low-potency synthetic cannabinoids can be prescribed for nausea associated with cancer chemotherapy and for the treatment of anorexia associated with weight loss in AIDS patients (Black, et al., 2019) (Food and Drug Administration, 2019). Cannabis-derived products have complex pharmacological characteristics, such as the opposing pharmacological and behavioral effects of the two main constituents of cannabis (Morgan, Schafer, Freeman, & Curran, 2010), further complicated by the high variability of cannabis products in terms of their pharmacodynamics and kinetics characteristics, as well as their delivery through various routes of administration.

In addition to the approved indications for which there is a scientific evidence, there is a hype on the use of many cannabis-derived products for a variety of conditions (Eisenstein, 2019; Stith, Vigil, Brockelman, Keeling, & Hall, 2019), with the market of CBD in the wellness industry in the US predicted to be at $24 billion by 2023 (Giammona C & Einhorn, 2019). The previous history of cannabis as a recreational drug could also potentially lead to a polarized

view on its medical use, with either an uncritical support independent on scientific evidence or a negative bias. In this context, the public will often gather non-specialist information on the Web rather than seeking advice from their doctor.

We have observed that often information online does not match scientific evidence, potentially pointing the public to use of health supplements for indications for which there is not high quality evidence, as in the case of probiotics or antioxidant supplements (Aslam, Gibbons, & Ghezzi, 2017; Neunez, Goldman, & Ghezzi, 2020). Cochrane reviews also showed that cannabinoids use is associated with an increased risk of transient adverse events including weakness, dizziness, sleepiness, difficulty with concentration, memory loss, confusion, headache, nausea and fatigue (Kafil, Nguyen, MacDonald, & Chande, 2018a, 2018b; Smith, Azariah, Lavender, Stoner, & Bettiol, 2015) and complete information should also describe these side effects.

This study aims at assessing the quality of the information available online on medical cannabis using a methodology used previously. We searched "medical cannabis" in google.com and analyzed the first 200 websites. We looked first at the typology of websites, whether professional, commercial, new or others, and their trustworthiness indicators such as the Journal of the American Medical Association (JAMA) (Silberg, Lundberg, & Musacchio, 1997) and the presence of the Health-On-the-Net (HON) certification (Boyer, Selby, Scherrer, & Appel, 1998). We then analyzed the content to find which disease or indications were mentioned by the web pages and correlated them with the number of clinical trials or reviews available in the Cochrane center. Finally, we considered whether information was complete and if webpages mentioned potential side effects and legal/regulatory issues

## 2. Methods

The search term "medical cannabis" was used as Google trends showed this as one of the top five topics related to worldwide searches on "cannabis". Searches were made on google.com from Brighton, UK in June 2019, using the URL google.com/ncr, to avoid redirection to the regional version of the search engine, and after deleting cookies and browsing history, to limit personalization of the search results. Based on previous studies, we aimed at collecting a sample of 150-170 webpages. We collected the first 244 links. Of these, 67 websites were excluded for the following reasons: Eight were referring to or selling a book; 17 were index pages, aggregators, or dynamic pages returning results from a search; 2 contained no information as they were just list of doctors or government offices; 14 were inaccessible or blocked websites or dead links; 17 contained no information about medical cannabis; 5 were about a new degree on medical cannabis opened at the University of Maryland; 3 were links to a video. Therefore, 176 webpages were analyzed for their content. The webpages were visited and analyzed using a previously validated methodology, based on intrinsic criteria and content. Analysis was done by four raters (ACM, AOVdF, IB and PG) and disagreements were resolved by discussion.

Intrinsic criteria:

1) Websites were classified as commercial (C), government (G), health portals (HP), news (N), non-profit (NP) or scientific journal (SJ) as described elsewhere (Aslam, et al., 2017; Neunez, et al., 2020). Websites not belonging to any of these typologies or where it was difficult to establish a typology were classified as "others" (O).

2) JAMA score. A score of 0 to 4 was assigned based on the presence of the following information: author, date, references to the source of information provided and ownership of the website (Silberg, et al., 1997). The presence of each of these criteria was counted as 1; therefore, the JAMA score ranged from 0 to 4.

3) HONcode. The HONcode certification was detected by the presence of a valid HONcode seal of approval on the webpage (Boyer, et al., 1998).

Content analysis:

1) Indication. We recorded the disease or biological process for which the use of medical cannabis was mentioned.

2) Stance about medical cannabis (in terms of efficacy or use), whether positive, neutral or negative. This was based on the wording of the text. Examples of classification of stance based on text contained in the webpage are shown in Supplementary Table 1

3) We recorded whether the webpage mentioned potential side effects and regulatory/legal issues associated with the use of cannabis products.

## 3. Results

### 3.1. Type of websites

Figure 1 shows the website typologies in the whole SERP (176 webpages) and in the top ten returned by Google. In the whole SERP, the most frequent typology was represented by websites from news outlets (52%) followed by government websites (14%). No news websites, however, were present in the top ten results, where a higher ranking was given by Google to websites from no-profit organizations (20% in the top ten vs. 8% in the whole SERP), health portals (20% vs. 4%) and government websites (30% vs. 14%). When comparing the frequency of each typology in the top ten results versus the rest of the SERP only news websites where significantly under-represented (Fisher's test followed by adjustment for eight multiple comparison using the two-stage linear step-up procedure of Benjamini, Krieger and Yekutieli set at a false discovery rate of 5%).

### 3.2. Trustworthiness indicators

Overall, the 176 pages had a median JAMA score of 3, interquartile range (IQR) 3,4, min 0, max 4. However, the JAMA score differed significantly across the different typologies of websites, with health portals and scientific journals scoring the highest, as shown in Figure 2. Commercial, government and non-profit webpages scored significantly lower when compared with the rest of the SERP (Mann-Whitney test followed by adjustment for eight multiple comparisons using the two-stage linear step-up procedure of Benjamini, Krieger and Yekutieli set at a false discovery rate of 5%).

Only 8 of the 176 websites had a HONcode certification, four of which were health portals, so that 57% of this type of websites were HON-certified. Three of the top ten websites (30%) had a HONcode, significantly more than the 3% (5/166) in the remaining websites (P=0.0064 by a two-tailed Fisher's test).

### 3.3. Content analysis. Disease and conditions mentioned.

As mentioned earlier, cannabis-derived products are approved for a limited number of indications. However, the webpages analyzed mentioned many more disease and conditions in relation to the possible benefits of medical cannabis. In this respect we wondered whether the frequency with which these conditions are ranked reflects, if not the approval by regulatory agencies, at least the amount of clinical research. As a proxy to the clinical research activity on medical cannabis we searched the Cochrane library on November 22$^{nd}$, 2019 and considered the numbers of clinical trials and Cochrane reviews in the database.

The results are shown in Figure 3. It can be seen that the indications most frequently mentioned by webpages are pain, epilepsy and multiple sclerosis. It is interesting that depression and psychosis/schizophrenia are among the least mentioned despite the high research activity in terms of clinical trials. In general, webpages mention a large number of conditions for which medical cannabis could have benefits, far more than those indications for which these products have been approved by regulatory agencies.

There was a difference in the number of diseases mentioned across different type of websites. As shown in Figure 4, health portals mentioned the largest number of diseases (median, 13; IQR, 9,15) and news the lowest (median, 2; IQR 0,5). Both values were significantly different from the rest of the SERP (Mann-Whitney test followed by adjustment for eight multiple comparisons using the two-stage linear step-up procedure of Benjamini, Krieger and Yekutieli set at a false discovery rate of 5%).

### 3.4. Completeness of information

We also assessed whether webpages report potential side effects and legal/regulatory issues of medical cannabis products. Side effects were mentioned by only 22% of webpages, legal/regulatory aspects by 80% of them. However, there were differences in the completeness of the information provided across typologies, particularly for side effects. As shown in Table 1, all health portals mentioned the side effects of medical cannabis, a frequency that was significantly higher when compared with the remaining 169 webpages; side effects were also more frequently mentioned by websites from non-profit organizations. On the contrary, only 5 of the 91 news websites mentioned side effects, significantly less than the remaining 85 webpages in the search. There were no significant differences in the frequency of mention of regulatory aspects.

### 3.5. Stance towards medical cannabis

The majority (81%) of the webpages had a stance towards medical cannabis which we defined as "neutral", 16% clearly positive about its benefits and only 2% negative. A sub-analysis in Figure 5 shows differences among the typologies of websites. The highest proportion of positive pages (67%) was observed in commercial websites, followed by non-profit; both frequencies were significantly higher when compared with the rest of the SERP (two-tailed Fisher's test followed by adjustment for eight multiple comparisons using the two-stage linear step-up procedure of Benjamini, Krieger and Yekutieli set at a false discovery rate of 5%.). Government and news websites had the lowest frequency of pages with a positive stance, which was significantly different only in the case of news. All pages from health portals and 88% of the news websites had a neutral view. Only four webpages had a negative stance on medical cannabis.

### 4. Discussion

This study shows that over half of the webpages containing information of medical cannabis are from news websites, which shows the newsworthiness of this topic. It should be mentioned, however, that the ranking made by Google prioritizes government websites, those from non-profit organizations and health portals over news outlets. Websites bearing the HONcode, an independent health information quality certification, were also ranked significantly higher, in agreement with our previous findings on information online on probiotics (Neunez, et al., 2020), confirming the observation that Google uses effective criteria to prioritize high-quality information.

Content analysis in terms of disease/indications showed that there is a mismatch between the therapeutic areas mentioned in the Web and those for which there are indications approved by regulatory agencies. The therapeutic area most frequently mentioned on the Web is pain. In the UK, NICE guidelines specifically recommend not to prescribe cannabis products for chronic pain unless as part of a clinical trial (NICE, 2019) and pain treatment is not an approved indication for any cannabis product in the US as of February 2020 (FDA, 2020).

The second most frequent indication mentioned in the Web is epilepsy. It should be noted, however, that the only cannabis-based medicine approved for this indication, plant-derived CBD, is licensed exclusively for the treatment of specific epileptic syndromes, namely Dravet and Lennox-Gastaut Syndromes (Friedman & Devinsky, 2015).

Another indication frequently mentioned is multiple sclerosis, despite the limited approval of cannabis-derived products in this disease. So far, an oro-mucosal spray containing plant-derived THC and CBD (nabiximols) is only approved, for instance in the UK, for the treatment of multiple sclerosis-associated spasticity.

A number of webpages mention cancer and chemotherapy. In fact, there is moderate evidence that THC, dronabinol and nabilone may be useful for treating refractory chemotherapy-induced nausea and vomiting, although they failed to show superiority as compared to conventional drugs, particularly prochlorperazine (Smith, et al., 2015). In patients with HIV/AIDS, the studies have been of short duration and limited to a small number of patients, preventing any solid conclusion of efficacy (Lutge, Gray, & Siegfried, 2013).

Finally, scarce evidence suggest the CBD might be useful for the treatment of other neurological and psychiatric disorders, such as schizophrenia and anxiety (Black, et al., 2019).

Although there is little clinical evidence for these indications (Crippa, Guimaraes, Campos, & Zuardi, 2018; Rohleder, Muller, Lange, & Leweke, 2016), anxiety is mentioned by a significant number of webpages.

Therefore, the general picture is that there is a partial mismatch between the indications mentioned on the Web for cannabis-based products and the regulatory approval, particularly for the treatment of pain. This breadth of online information might potentially raise the interest of the public in the use of medical cannabis for a range of indications that is broader than that of indications that are actually approved, potentially making more attractive for the public to use cannabis supplements as self-medication in absence of a medical prescription.

On the other hand, we found that the large majority of websites (88%) had a neutral stance on the use of cannabis, indicating that the information available online is not particularly polarized. One exception is represented by commercial websites that have a largely positive stance (67%), which can be explained by the fact that commercial websites are often selling cannabis-derived products.

Another aspect of health information online is that of completeness of the information provided, that was suggested as a criterion for health information quality (Dutta-Bergman, 2004). In two recent studies we noted that a key aspect of completeness of health information is the mention of potential side effects and regulatory issues (Manley & Ghezzi, 2018; Neunez, et al., 2020). In the present study, while regulatory issues were often mentioned, only 22% of websites mentioned the potential side effects, and this was due, in particular, to the paucity of this type of information in news outlets and commercial websites. The websites providing the most complete information were those from health portals, all of them mentioning the potential adverse effects.

The main limitation of the present study is the protocol through which the sample of webpages were collected, as it may vary with time and with the search query. Although we used generic and neutral search terms, the results could be different when searching for cannabis and a specific health condition. This limitation, however, could have been at least partially circumvented by evaluating a large number of webpages. Another limitation is that we only used Google as a search engine. These may not be a representative sample of the infosphere, because Google recently implemented high quality standards in the page ranking of what they define "your money your life" pages. These standards rank higher pages that are written by people or organizations with medical expertise, authoritativeness and trustworthiness, and where the information provided is aligned with the scientific consensus on the topic https://static.googleusercontent.com/media/guidelines.raterhub.com/en//searchqualityevaluatorguidelines.pdf). As we noted elsewhere, other search engines provide more lower-quality results than Google (https://arxiv.org/abs/1912.00898). On the other hand, as Google has around 90% of the search engine market share, the websites it returns are the ones that the user would likely find.

In conclusion, our study indicates that the information available on the Web can raise unrealistic expectations in the public and contribute to a hype that could potentially lead patients to use cannabis-based products as a self-medication for indications for which there is no strong evidence of efficacy and with an unclear safety profile.

Table 1. Mention of side effects and legal/regulatory aspects across different website typologies.

| Typology | Side effects | Legal/regulatory |
|---|---|---|
| C | 17% (1/6) | 50% (3/6) |
| G | 25% (6/24) | 92% (22/24) |
| HP | 100% (7/7)* | 71% (5/7) |
| N | 5% (5/91)* | 84% (76/91) |
| NP | 57% (8/14)* | 71% (10/14) |
| P | 44% (4/9) | 67% (6/9) |
| SJ | 33% (4/12) | 58% (7/12) |
| O | 31% (4/13) | 85% (11/13) |

Significantly different vs the rest of the SERP by a two-tailed Fisher's test followed by adjustment for eight multiple comparisons using the two-stage linear step-up procedure of Benjamini, Krieger and Yekutieli set at a false discovery rate of 5%.

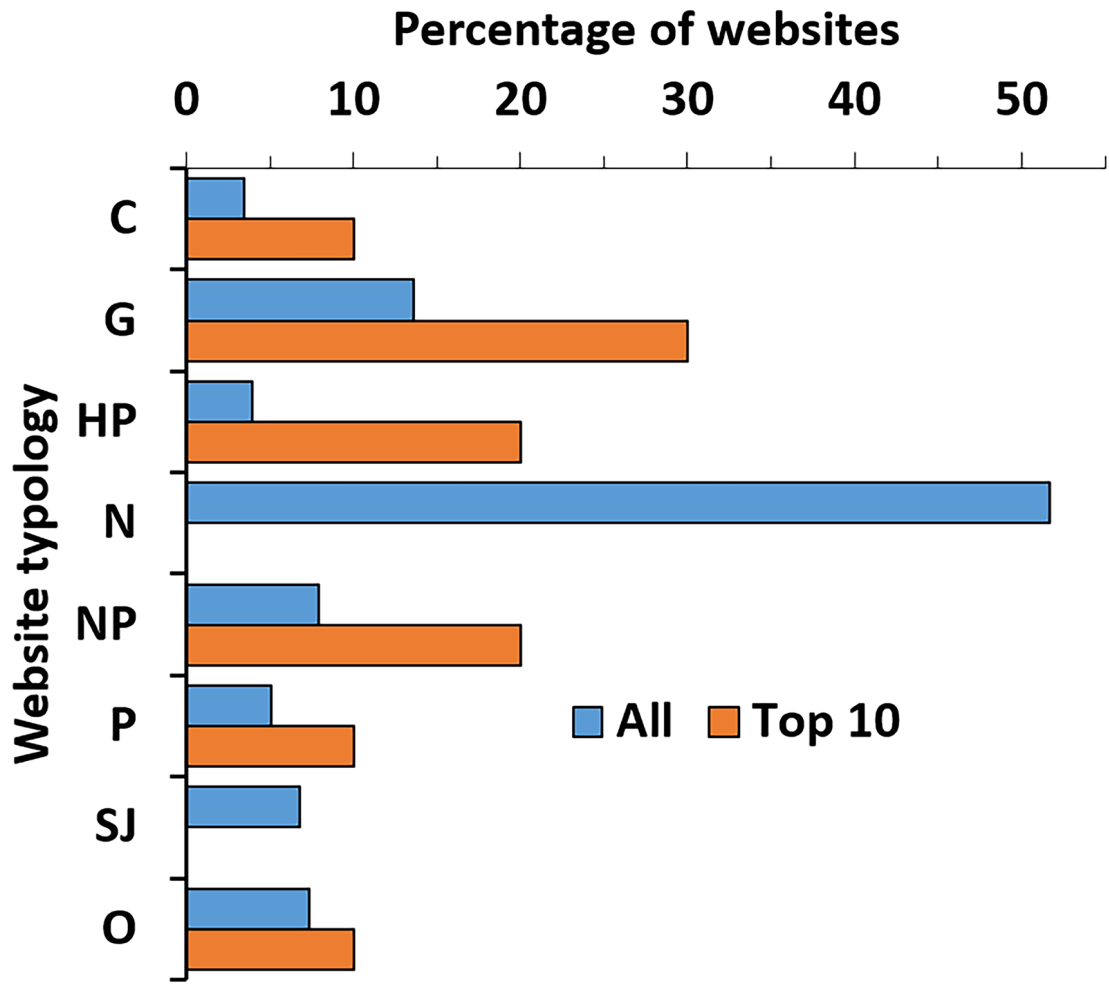

Fig.1. Typologies of websites returned by Google.

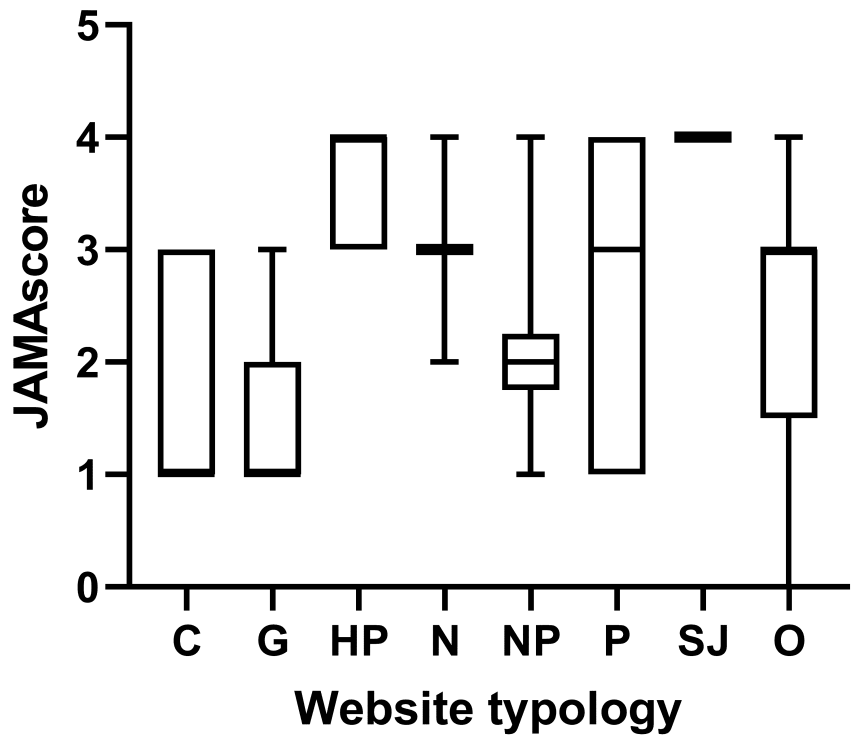

Fig 2. JAMA score of webpages of different typologies. Data represent the median, IQR, minimum and maximum. Legend: C, commercial; G, government; HP, health portals; N, news, NP, no-profit; P, professional; SJ, scientific journals; O, others (unclassified).

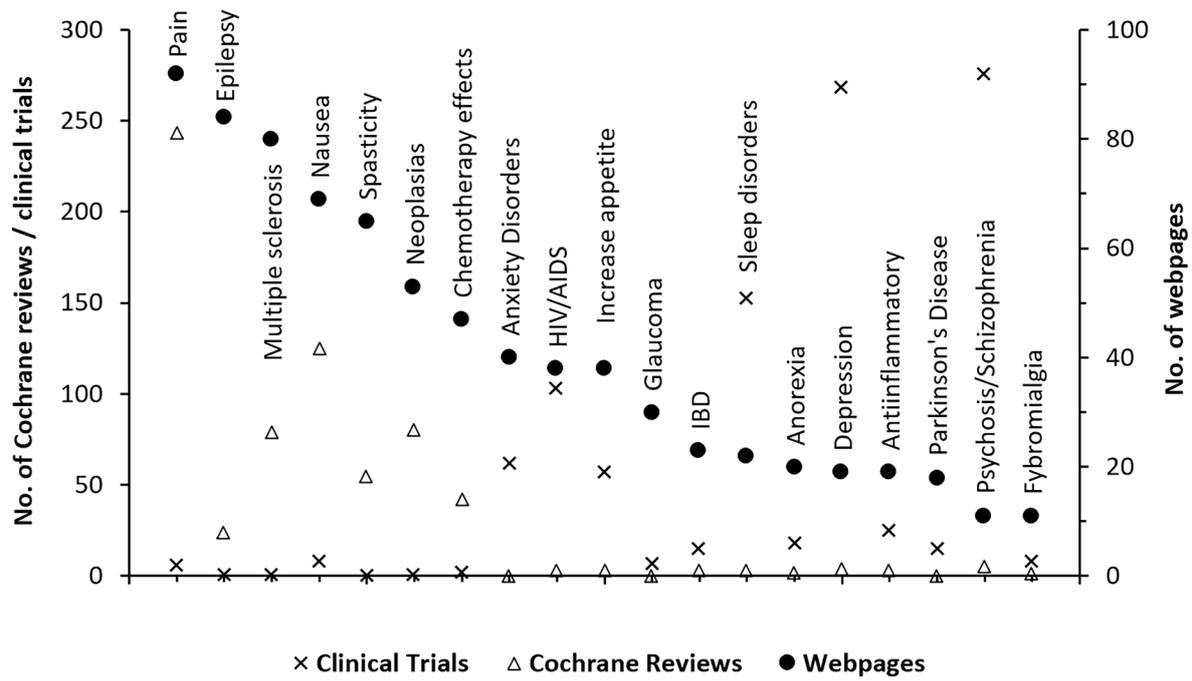

Fig. 3. Therapeutic areas mentioned in the webpage (solid circles) in relation to the number of Cochrane reviews (triangles) and clinical trials (x) in the Cochrane database for that indication

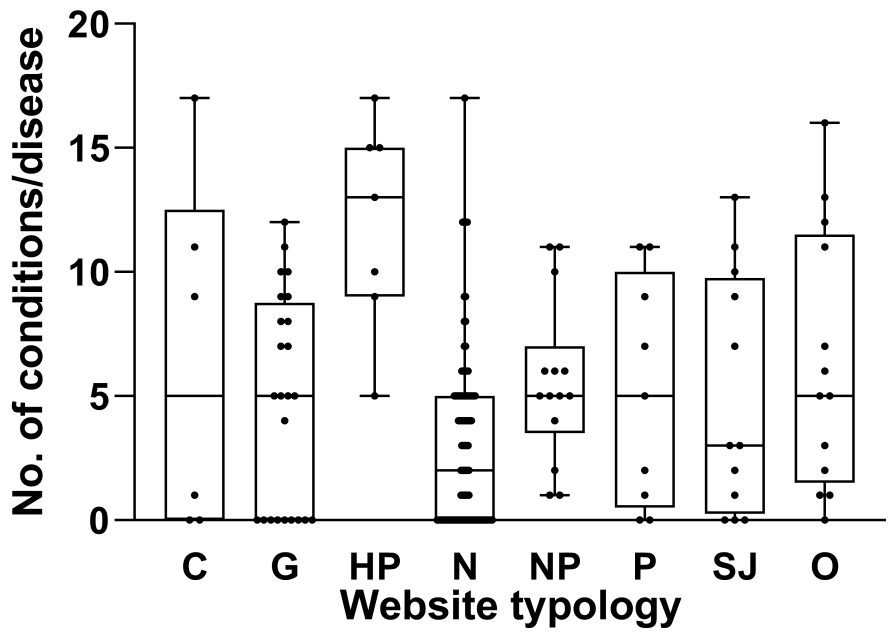

Fig. 4. Number of indications mentioned by webpages of different typologies. Data are median, IQR, minimum, maximum.

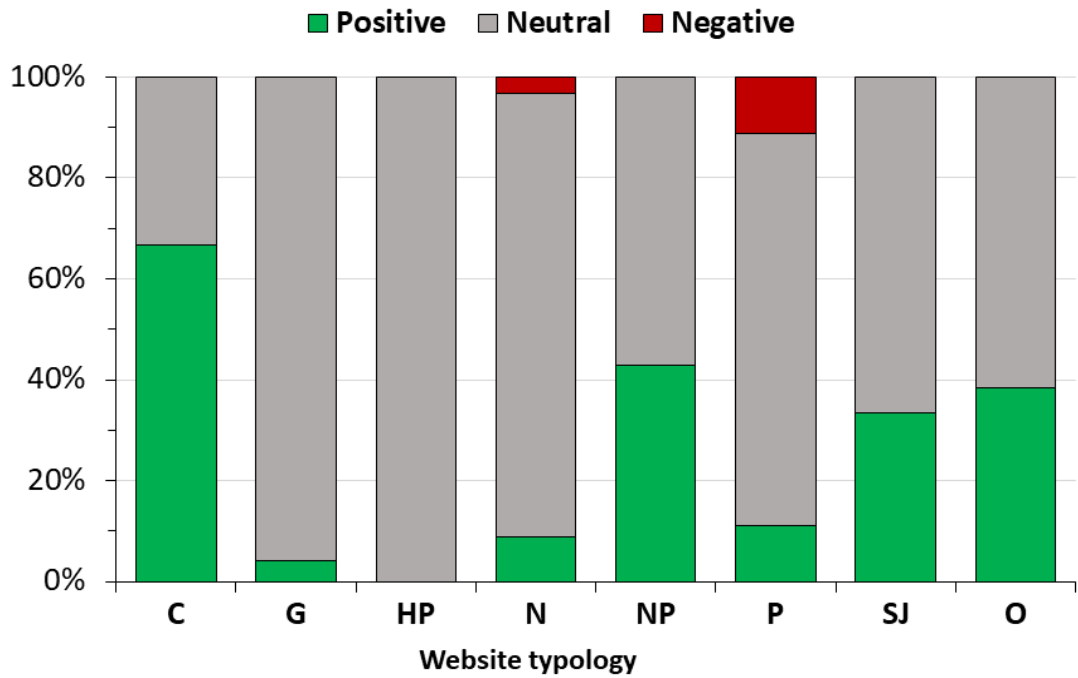

Fig. 5. Stance toward medical cannabis in different typologies of websites

**Supplementary Table 1. Examples of stance about cannabis**

| |
|---|
| *Examples of positive webpages* |
| "Marijuana has been shown to alleviate symptoms of a huge variety of serious medical conditions including cancer, AIDS, and glaucoma, and is often an effective alternative to synthetic painkillers" (http://www.drugpolicy.org/issues/medical-marijuana) |
| "Both forms of cannabis, hemp and marijuana, have been shown to contain medically beneficial levels of differing cannabinoids, active compounds found in the cannabis plant" (https://www.medicalmarijuanainc.com/) |
| "Our study suggest that Cannabis therapy, as an adjunct a traditional analgesic therapy, can be an efficacious tool to make more effective the management of chronic pain and its consequences on functional and psychological dimension. Further randomized, controlled trials are needed to confirm our conclusions" (https://www.ncbi.nlm.nih.gov/pubmed/29938740) |
| "The claims by medical users that cannabis reduces the symptoms of MS has been confirmed by UK government trials" ( https://www.ukcia.org/medical/index.php) |
| "However, reviews of published studies have generally shown that synthetic cannabinoids favorably impact symptoms of pain and spasticity" (https://www.nationalmssociety.org/Treating-MS/Complementary-Alternative-Medicines/Marijuana/Marijuana-FAQs) |
| *Examples of negative webpages* |
| "According to the Department of Health, it is important to note that where a cannabis product is a specified controlled drug legally permitted for medical use, in connection with the MCAP, this does not signify any endorsement whatsoever of the safety, quality or efficacy of the specified controlled drug for the indication prescribed" (https://www.imt.ie/news/minister-no-liability-access-medical-cannabis-programme-26-06-2019/) |
| "Medical marijuana dispensing is associated with reduced perception of marijuana-related risks and increased rates of marijuana use among adolescents" (https://www.aacap.org/AACAP/Policy_Statements/2012/AACAP_Medical_Marijuana_Policy_Statement.aspx) |
| *Examples of neutral webpages* |
| "While cannabis, or marijuana, has been around for a long time, there is still not much formal evidence for doctors to rely on if they are thinking about prescribing a medicinal cannabis product to a patient. The Commonwealth Government has released documents which summarise the evidence so far that medicinal cannabis may be useful in treating some conditions "(https://www.betterhealth.vic.gov.au/health/conditionsandtreatments/medicinal-cannabis) |
| "The use of cannabis as medicine has not been rigorously tested due to production and governmental restrictions, resulting in limited clinical research to define the safety and efficacy of using cannabis to treat diseases. Preliminary evidence suggests that cannabis can reduce nausea and vomiting during chemotherapy, improve appetite in people with HIV/AIDS, reduces chronic pain and muscle spasms and treats severe forms of epilepsy" (https://en.wikipedia.org/wiki/Medical_cannabis) |
| "So far, researchers haven't conducted enough large-scale clinical trials that show that the benefits of the marijuana plant (as opposed to its cannabinoid ingredients) outweigh its risks in patients it's meant to treat" https://www.drugabuse.gov/publications/drugfacts/marijuana-medicine) |
| "There haven't been any studies comparing medical cannabis with other medicines already licensed for treating epilepsy. So we don't know if medical cannabis is more or less effective than other epilepsy treatments. Neither do we know if it is more or less safe than other epilepsy treatments" (https://www.epilepsy.org.uk/info/treatment/cannabis-based-treatments) |